\documentclass[fleqn,usenatbib,useAMS]{mnras}

\usepackage{graphicx}	
\usepackage{amsmath}	
\usepackage{amssymb}	
\usepackage{multicol}        
\usepackage{bm}		
\usepackage{pdflscape}	
\usepackage[T1]{fontenc}
\usepackage{ae,aecompl}

\numberwithin{equation}{section}

\title[Bridges and gaps at first-order resonances]{Bridges and gaps at low-eccentricity first-order resonances}

\author[K. I. Antoniadou and A.-S. Libert]{Kyriaki I. Antoniadou$^{1}$\thanks{Contact e-mail: kyant@auth.gr (KIA)}%
and Anne-Sophie Libert$^{2}$%
\\
$^{1}$Department of Physics, Aristotle University of Thessaloniki, 54124, Thessaloniki, Greece\\
$^{2}$naXys, Department of Mathematics, University of Namur, 61 Rue de Bruxelles, 5000 Namur, Belgium
}

\date{Last updated XXXX; in original form YYYY}

\pubyear{2021}

\begin{document}
\label{firstpage}
\pagerange{\pageref{firstpage}--\pageref{lastpage}}
\maketitle

\begin{abstract}
Previous works on the divergence of first-order mean-motion resonances (MMRs) have studied in detail the extent of the pericentric and apocentric libration zones of adjacent first-order MMRs, highlighting possible bridges between them in the low-eccentricity circular restricted three-body problem. Here, we describe the previous results in the context of periodic orbits and show that the so-called circular family of periodic orbits is the path that can drive the passage between neighbouring resonances under dissipative effects. We illustrate that the circular family can bridge first and higher order resonances while its gaps at first-order MMRs can serve as boundaries that stop transitions between resonances. In particular, for the Sun-asteroid-Jupiter problem, we show that, during the migration of Jupiter in the protoplanetary disc, a system initially evolving below the apocentric branch of a first-order MMR follows the circular family and can either be captured into the pericentric branch of an adjacent first-order MMR if the orbital migration is rapid or in a higher order MMR in case of slow migration. Radial transport via the circular family can be extended to many small body and planetary system configurations undergoing dissipative effects (e.g., tidal dissipation, solar mass-loss and gas drag).
\end{abstract}

\begin{keywords}
celestial mechanics --  chaos -- minor planets, asteroids: general -- planets and satellites: dynamical evolution and stability 
\end{keywords}



\section{Introduction}
Obtaining an explicit view of the phase space of an underlying dynamical system provides important information about its long-term evolution and stability. The periodic orbits play a crucial role in shaping the regular and chaotic domains. More particularly, the periodic orbits correspond to fixed or periodic points on a given Poincar\'e map in a specific rotating frame of reference \citep{pnc}, or equivalently, they correspond to the stationary solutions of an averaged Hamiltonian, as long as it is adequately accurate (high-order truncation of the series expansion of the disturbing function), which are also called apsidal corotation resonances (e.g. in \citet{beau03}). The periodic orbits showcase the exact location of a mean-motion resonance (MMR) in phase space. It is widely known in Hamiltonian systems that invariant tori exist around the stable periodic orbits in phase space, whereas homoclinic webs are formed around the unstable ones \citep{arnold}. In the former case, the motion is quasi-periodic and the resonant angles librate, with the libration to be taking place about the exact periodic orbit, while in the latter case, the motion is chaotic and the resonant angles rotate. In planetary systems, the stable periodic orbits drive the migration process of planets to orbitally stable resonant configurations \citep[see e.g.][]{bmfm06,hadjvoy10,vat14,libsa18}. While the planar circular restricted three-body problem (CRTBP) has been extensively used as a model for the study of the motion of asteroids in MMR with Jupiter, the orbital migration via the planar families of the CRTBP has not yet been deeply explored for prograde orbits. 

Recently, \citet{malzha} undertook an exploration of low ($e\rightarrow 0$)- up to high ($e\rightarrow 1$)-eccentricity prograde orbits of the planar CRTBP with Poincar\'e sections as a means of measuring the libration widths in the 4/3, 3/2 and 2/1 first-order MMRs. The system studied was a massless body (asteroid) moving in an interior orbit to Jupiter, while both of them are locked in an MMR. They illustrated how the centres of the pericentric and apocentric resonance zones diverge away from the nominal resonance at very low-eccentricity of the test particle. Also, they  showed how the pericentric and apocentric libration zones of adjacent first-order MMRs seem to connect, which could constitute possible ``bridges'' for radial transport in planetary systems. These results were reproduced with an analytical Hamiltonian approach by \citet{leili20}, who also extended the study of the libration widths (based on an analysis of phase portraits) to first-order exterior MMRs. Later on, \citet{hanlun} also studied retrograde orbits and coorbital motion.

\begin{figure*}
\includegraphics[width=16cm]{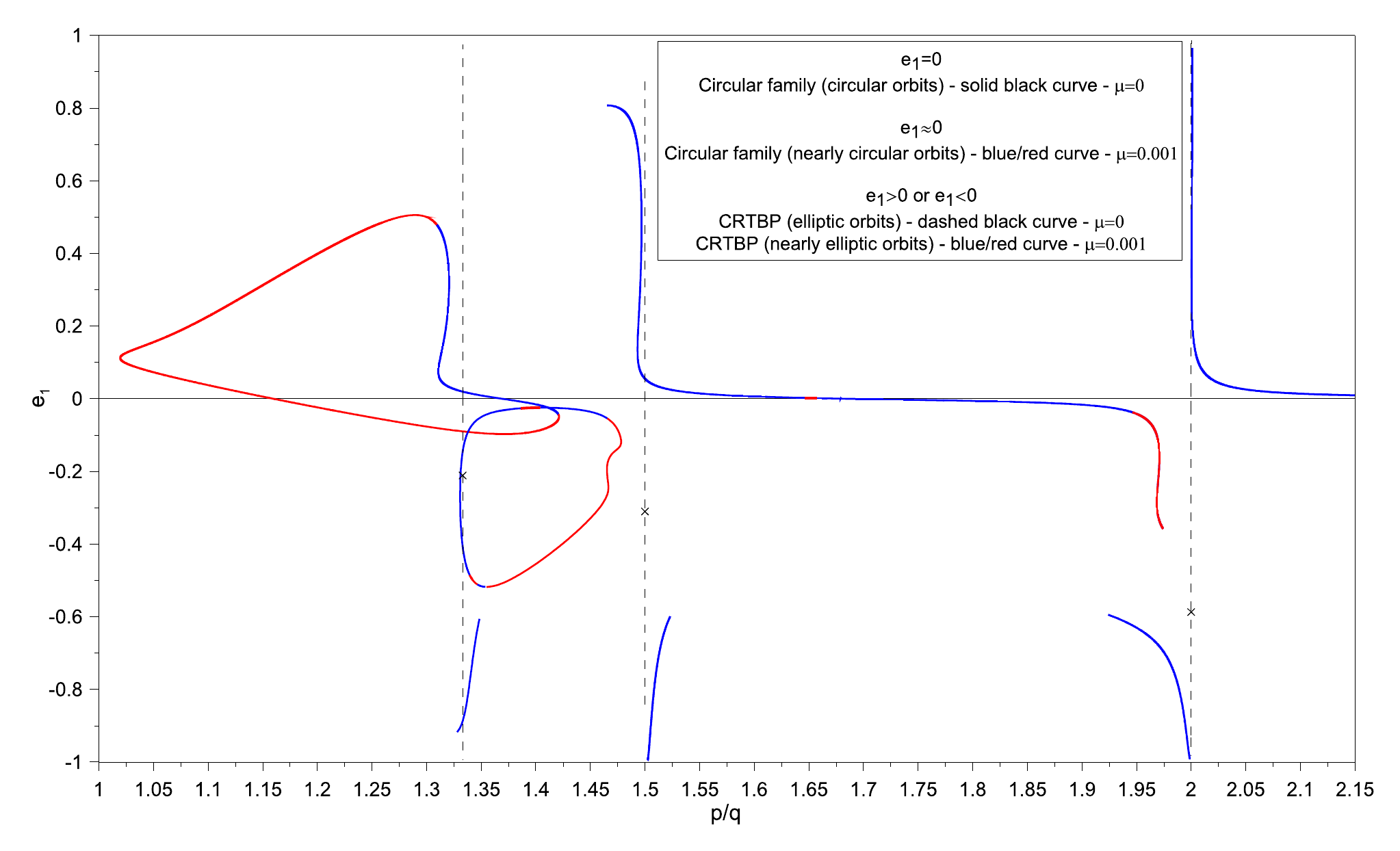}
\caption{Bridges and gaps at low-eccentricity first-order resonant orbits. For the perturbed case ($\mu=0.001$), we present the circular family (coloured curves at $e_1\approx 0$) and the families of periodic orbits of first-order MMRs in the CRTBP (coloured curves for $e_1>0$). Hyperbolic deviation is clearly visible (gaps) at 4/3, 3/2 and 2/1 MMRs and the pericentric (apocentric) branch is indicated by positive (negative) values of $e_1$. For the first-order MMRs, the resonant angle $\theta_{1,p/q}$ equals to $0^{\circ}$ at the pericentric branches and $180^{\circ}$ at the apocentric ones. Blue (red) colour stands for linearly stable (unstable) orbits. The highly eccentric stable orbits of the apocentric branches located after the close encounters (symbol ``x'') are also shown. For the unperturbed case ($\mu=0$), the circular family located at $e_1=0$ is depicted with a solid black line and the two branches of the families of periodic orbits in the CRTBP with dashed black lines.}
\label{fig1}
\end{figure*}

In this work, we aim to describe those previous results on the prograde interior resonances in light of the existing periodic orbits that dominate the dynamics observed in the CRTBP and showcase possible connections between first and higher order MMRs under the effect of dissipative forces. In Sect.~\ref{reality}, we put the previous findings in the context of periodic orbits. In particular, we highlight the importance of the so-called circular family of periodic orbits as well as the bridges and gaps observed along the family. For the sake of completeness, a technical review on the origin, continuation, and bifurcation of (prograde and retrograde) periodic orbits starting from the circular family and moving on to the families of periodic orbits of the CRTBP is given in Appendix~\ref{appendix}. In Sect.~\ref{mig}, we investigate the possibility of passages between first-order MMRs. The circular family and the families in the CRTBP are used as paths that drive the orbital migration of asteroids captured in MMR with Jupiter to a dynamically stable end-up state. In Sect.~\ref{con}, we summarize our findings on transitions between MMRs and resonance captures along these paths.

\section{Bridges and gaps in the context of periodic orbits}\label{reality}
We consider two primaries, the Sun with mass equal to $1-\mu$ and Jupiter on a circular orbit with mass equal to $\mu=0.001$, as well as an asteroid in an interior orbit to Jupiter as the massless body in the planar CRTBP \citep[see e.g.][]{sze,roy}. In the following, subscripts 1 and 2 will refer to the asteroid and Jupiter, respectively. 

The continuation method, which is used to prove the existence of periodic orbits, is based on the variation of a parameter, given a very small perturbation of the initial system, e.g. the mass or the period \citep{aren66,siemos,aren76}. This method results in the analytical computation of smooth curves, which are also called characteristic curves or \emph{families} in literature. A thorough review of the families of periodic orbits in the CRTBP is provided in Appendix~\ref{appendix}, for both the unperturbed case ($\mu=0$) and perturbed case ($\mu=0.001$), in the context of direct (prograde) and retrograde orbits. In this section, we focus on the low-eccentricity first-order resonance direct orbits with the aim of putting into perspective the recent studies of \citet{malzha} and \citet{leili20}.

We consider the \textit{circular family} \cite[see e.g.][]{col68,hadjich84} consisting of nearly circular resonant and non-resonant orbits of the asteroid (coloured blue/red curves of Fig.~\ref{fig1} at $e_1\approx 0$)\footnote{When $\mu=0$, both eccentricities are exactly equal to 0 and the circular family is showcased in Fig.~\ref{fig1} by the solid black line at $e_1=0$. We refer to Appendix~\ref{appendix} for more details on the differences between the cases of $\mu=0$ (Appendix~\ref{unpert}) and $\mu>0$ (Appendix~\ref{pert}).}. Along the circular family, the resonance $p/q$ (or equivalently the period or the semimajor axis of the asteroid) varies. In Fig.~\ref{fig1}, the portion in between the 4/3 and 2/1 MMRs is shown. At rational values of the resonance along the circular family, we have the generation of two new \textit{branches} or \textit{families} in the CRTBP consisting of nearly elliptic ($e_1>0$) periodic orbits. Along these branches, the resonance remains almost constant and the eccentricity of the asteroid varies and can take values $e_1\rightarrow 1$. One branch represents the location of the asteroid at pericentre and the other at its location at apocentre. In Fig.~\ref{fig1}, these branches\footnote{The branches associated with $\mu=0$ are depicted with dashed black lines, see Appendix~\ref{unpert}.} are depicted with coloured curves for the 4/3, 3/2 and 2/1 MMRs, with positive and negative eccentricity values depending on the pericentric/apocentric phase of the asteroid. In the following, we will use the resonant angle
\begin{equation} \label{angle}
\theta_{1,p/q}=q\lambda_1-p\lambda_2+(p-q)\varpi_1,
\end{equation}
where $\lambda_2=0^{\circ}$, since, without loss of generality, we take $\varpi_2=M_2=0^{\circ}$ as the outer body, Jupiter, always describes circular orbits. In particular, for the first-order resonant orbits presented in Fig. \ref{fig1}, we have $\theta_{1,p/q}=0^{\circ}$ at the pericentric branches of the CRTBP (positive eccentricity values) and $\theta_{1,p/q}=180^{\circ}$ at the apocentric branches (negative eccentricity values).

\citet{gui69,schmiB,schmi72,hadra} analytically proved that the first-order MMRs cannot be continued when $\mu>0$ and the circular family deviates hyperbolically and follows the family of the periodic orbits of the CRTBP. The hyperbolic deviation is clearly visible at 4/3, 3/2 and 2/1 MMRs in Fig.~\ref{fig1}. As a result, the continuous curve that existed for $\mu=0$ (solid black line at $e_1=0$) breaks when $\mu>0$ and {\it gaps} are formed\footnote{Note that for $\mu=0$ the dashed lines (pericentric and apocentric branches of the CRTBP) are smoothly connected with the same resonant orbit of the solid black line (circular family) at first-order MMRs.}. Therefore, first-order MMRs have two distinct libration zones, one for pericentric motion and one for apocentric motion (as it can also be deduced from the Poincar\'e sections of \citet{malzha}). Moreover, \citet{hadjich84} showed how the values of the MMRs along both branches vary at first-order MMRs, which means that the resonance is displaced slightly from its nominal value. 

Regarding the stability of the orbits, the circular family when $\mu>0$ possesses very small regions of instability only around second-order MMRs (small red segments for $e_1\approx 0$ at 5/3 and 7/5 MMRs in Fig.~\ref{fig1}) and is stable (blue) elsewhere. However, the two branches (pericentric and apocentric) of nearly elliptic orbits of the CRTBP differentiate: one branch is usually stable and the other unstable (see Appendix~\ref{pert} for details regarding the deduction of their linear stability). We note that the apocentric branches have highly eccentric stable periodic orbits, which are located after the close encounters between the asteroid and the planet (symbolized by ``x'' in Fig.~\ref{fig1}). In literature, the stability of the periodic orbits of the two branches of families in the CRTBP has widely been computed via the libration widths of resonant angles measured on Poincar\'e surface of sections, as in \citet{malzha}.

Therefore, as a figure of speech, the circular family acts like a {\it bridge} that connects orbits of adjacent resonances of different order. As previously said, second-order MMRs (like the 5/3 MMR in between the 2/1 and 3/2 MMRs or the 7/5 MMR in between the 3/2 and 4/3 MMRs for which a change of stability is observed) and also higher order MMRs are present along the circular family. Neighbouring first-order MMRs are then (broadly speaking) connected by bridges along the circular family if the second-order MMRs in between are surpassed. These possible bridges between the pericentric and apocentric libration zones of adjacent first-order MMRs were unveiled in theoretical studies on periodic orbits \citep[see e.g.][]{col68}. In the next section, we aim to assess their efficiency during the migration of small bodies and planets in the low-eccentricity regime. We provide several illustrations showing that the circular family can bridge first and higher order resonances while its gaps at first-order MMRs can serve as boundaries that stop transitions between resonances.

\section{Orbital migration and resonance captures}\label{mig}

\begin{figure}
\includegraphics[width=\columnwidth]{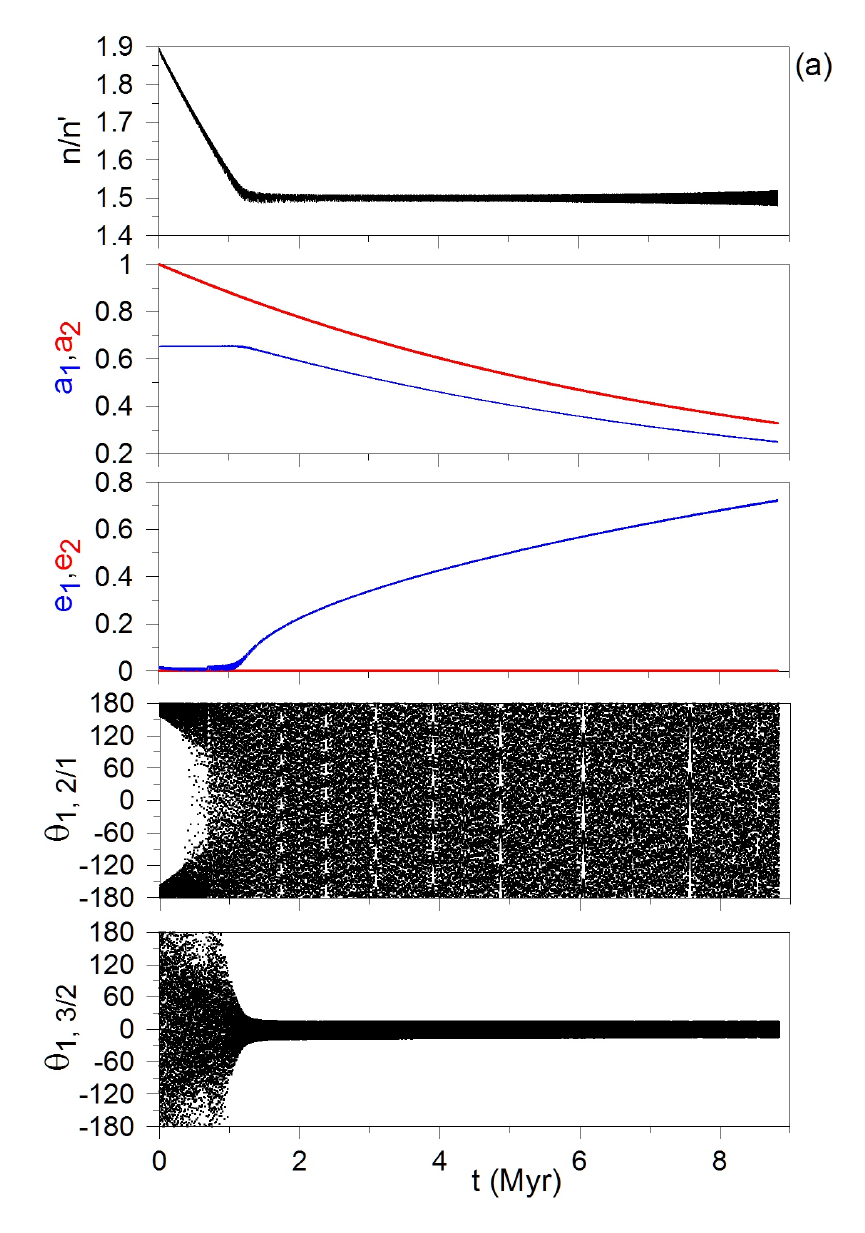}\vspace{-0.5cm}\\
\includegraphics[width=9cm]{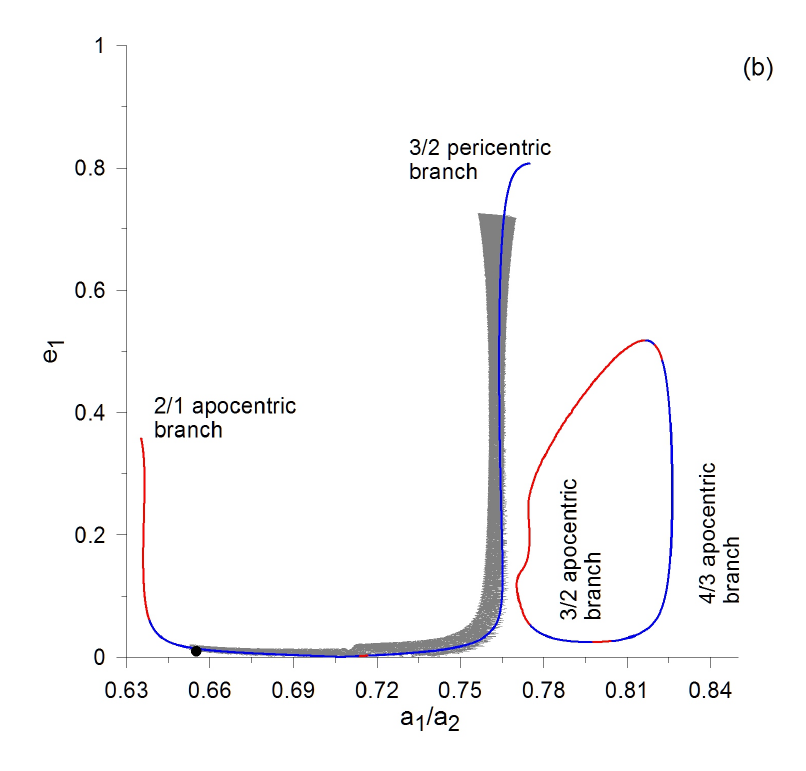}
\caption{Example of a transition from the 2/1 MMR to the 3/2 MMR. In panel (a), the evolution of the orbital elements and the resonant angles $\theta_{1,2/1}=\lambda_1-2\lambda_2+\varpi_1$ and $\theta_{1,3/2}=2\lambda_1-3\lambda_2+\varpi_1$ are shown. In panel (b), the evolution is overplotted with grey colour on the circular family ($e_1\approx 0$) and the families of the CRTBP ($e_1>0$). The initial condition is depicted by a black dot.}
\label{fig2}
\end{figure} 

\begin{figure}
\includegraphics[width=\columnwidth]{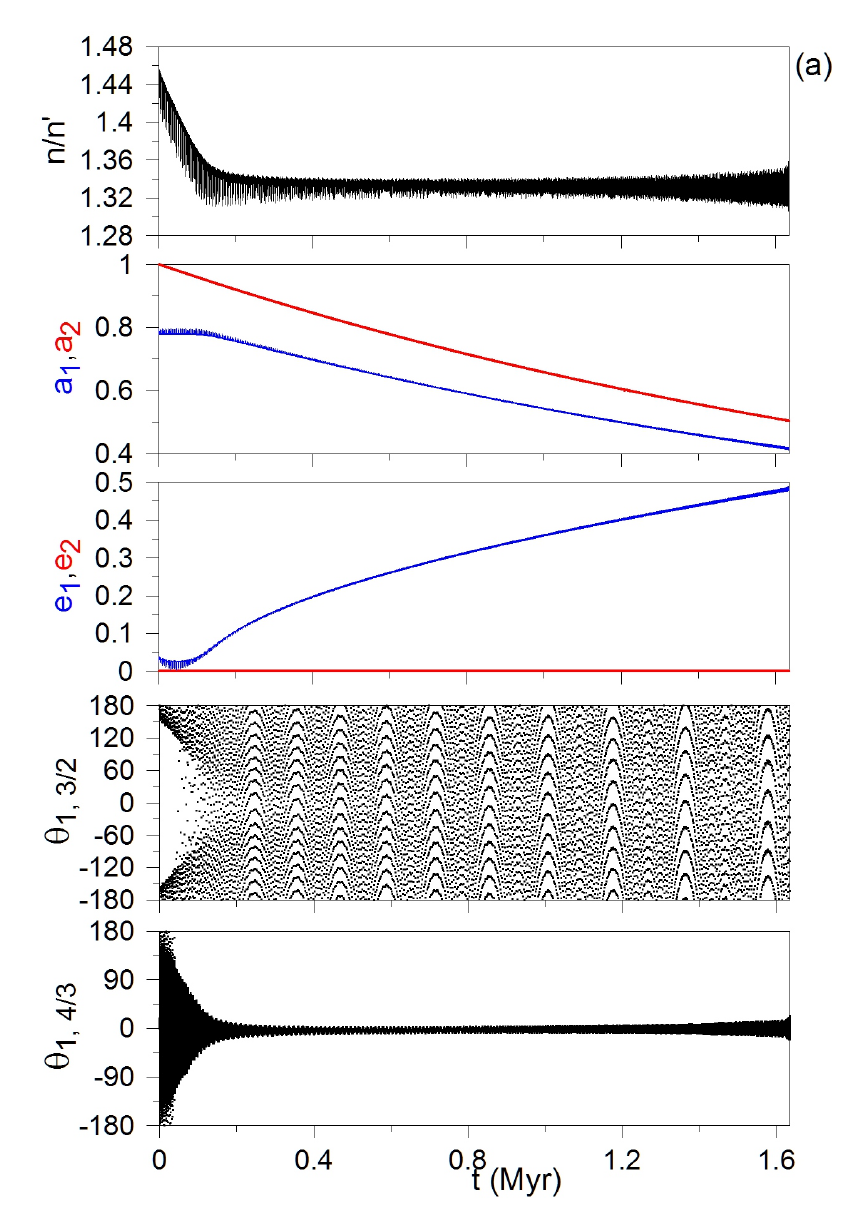}\vspace{-0.5cm}\\
\includegraphics[width=9cm]{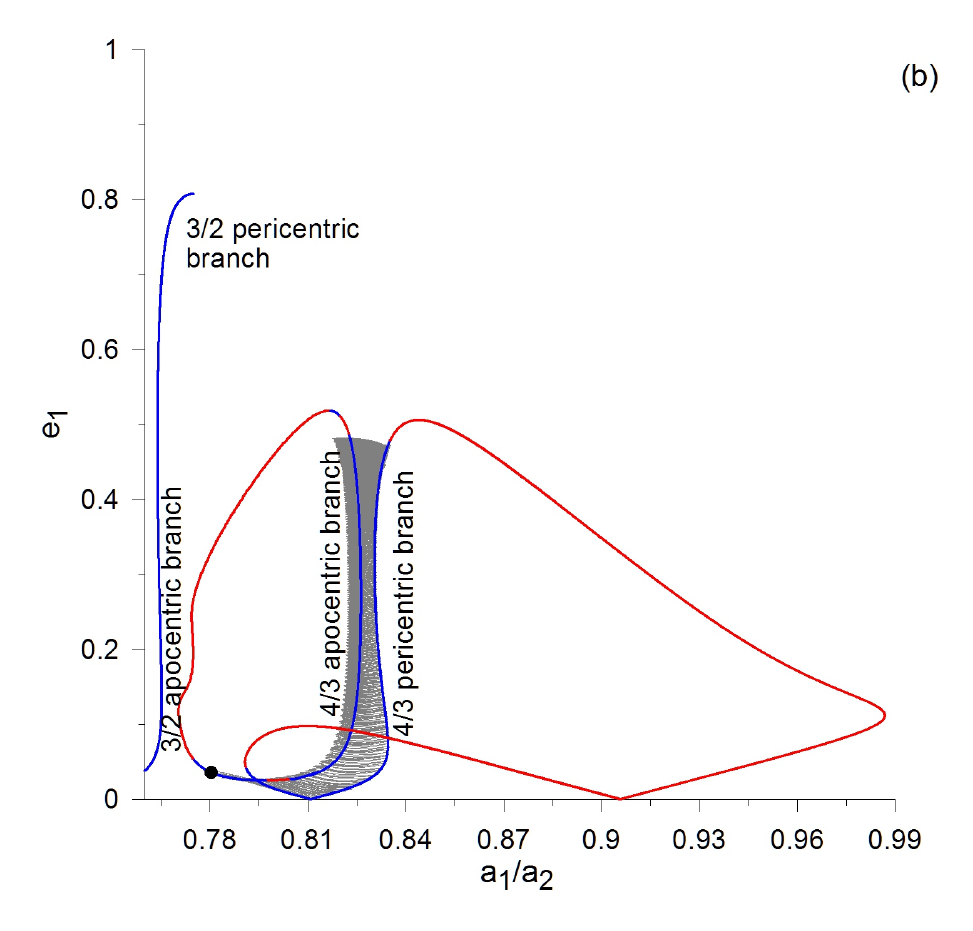}
\caption{A transition from the 3/2 MMR to the 4/3 MMR is shown and presented as in Fig. \ref{fig2} with $\theta_{1,4/3}=3\lambda_1-4\lambda_2+\varpi_1$. After the 4/3 resonance capture the evolution is delineated by the two branches of the 4/3 MMR in the CRTBP.}
\label{fig3}
\end{figure}

\begin{figure}
\includegraphics[width=\columnwidth]{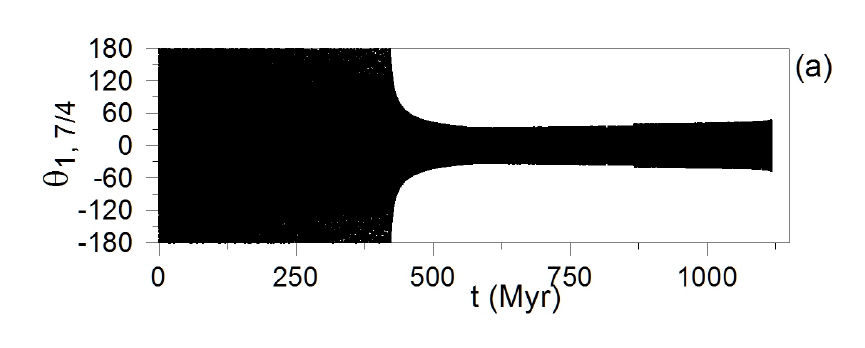}\vspace{-0.5cm}\\
\includegraphics[width=\columnwidth]{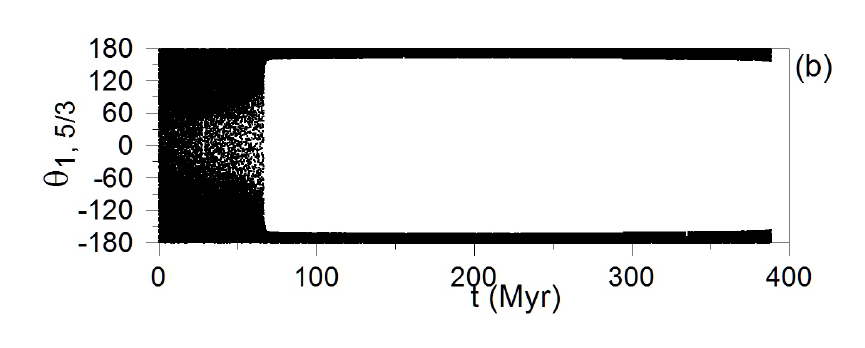}\\
\includegraphics[width=\columnwidth]{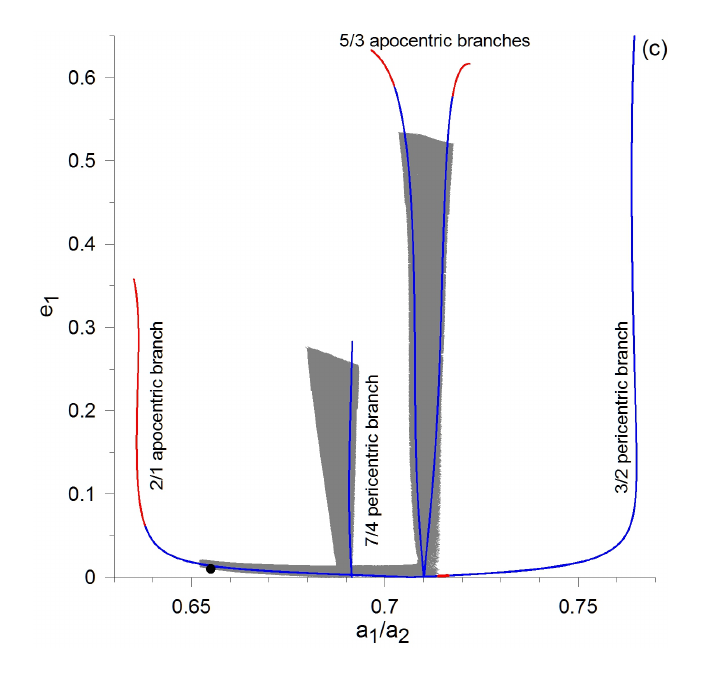}
\caption{Representative illustrations of transitions from the 2/1 MMR to high-order resonance captures. In panel (a), a capture in the 7/4 MMR (libration of $\theta_{1,7/4}=4\lambda_1-7\lambda_2+3\varpi_1$) is shown, while in panel (b), a capture in the 5/3 MMR (libration of $\theta_{1,5/3}=3\lambda_1-5\lambda_2+2\varpi_1$) is shown. These evolutions are overplotted with grey colour on the ($a_1/a_2,e_1$) plane in panel (c), along the families of 7/4 and 5/3 periodic orbits and more particularly, on the stable pericentric branch of the 7/4 MMR with $\theta_{1,7/4}=0^{\circ}$  and the stable apocentric branches of the 5/3 MMR with $\theta_{1,5/3}=180^{\circ}$.}
\label{fig4}
\end{figure}

\begin{figure}
\includegraphics[width=\columnwidth]{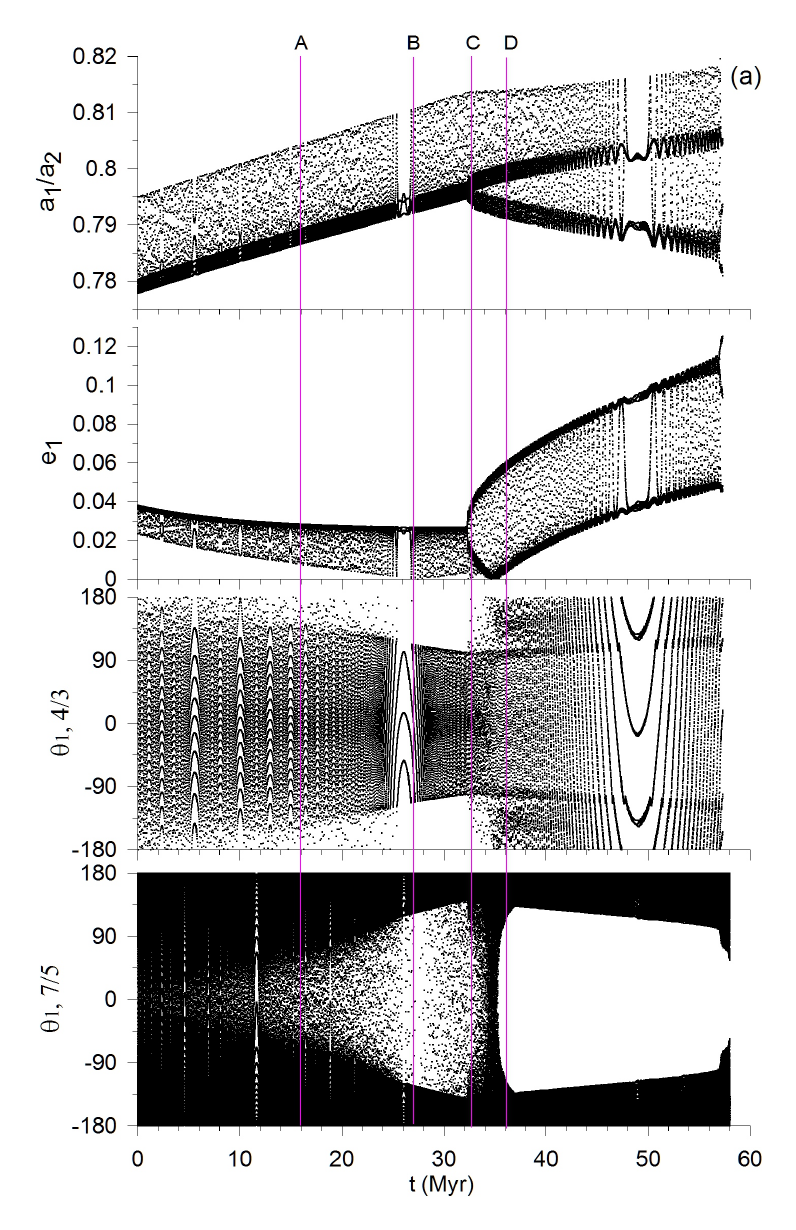}\vspace{-0.5cm} \\
\includegraphics[width=\columnwidth]{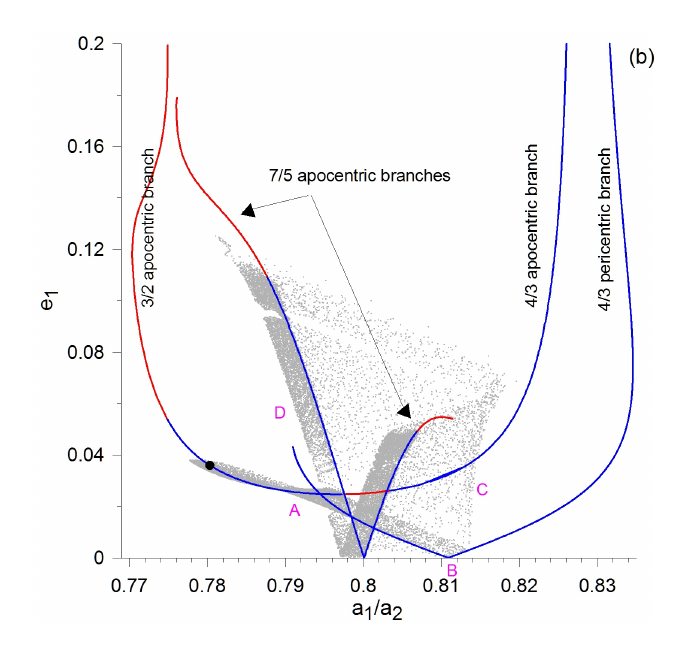}\vspace{-0.5cm}\\
\caption{A transition from the 3/2 MMR to the 7/5 MMR. In panel (a), the evolution of the orbital elements and the resonant angles $\theta_{1,4/3}$ and $\theta_{1,7/5}=5\lambda_1-7\lambda_2+2\varpi_1$ are shown. In panel (b), the evolution of the system is overplotted on the 7/5 stable apocentric branches with $\theta_{1,7/5}=180^{\circ}$. Labels A-D demonstrate different time domains and the path being followed during the evolution.}
\label{fig5}
\end{figure}

\begin{figure*}
\includegraphics[width=\textwidth]{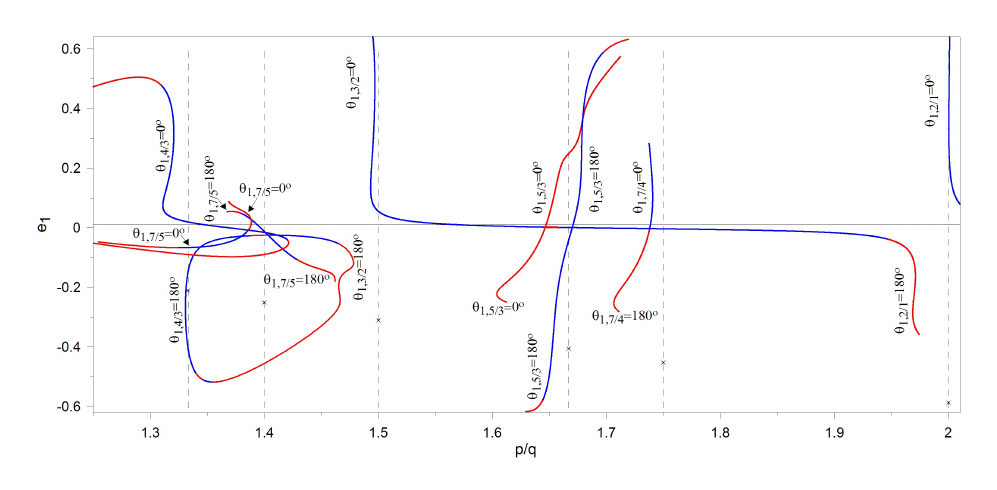}  
\caption{The families of 7/4, 5/3 and 7/5 resonant periodic orbits in the CRTBP, presented as in Fig. \ref{fig1}.}
\label{fig6}
\end{figure*}

We illustrate possible resonance captures along the circular family, which serves as a bridge between the gaps at first-order resonant orbits, in the context of the Sun-asteroid-Jupiter problem. In particular, we aim to see whether radial transport between adjacent first-order MMRs is possible under the effects of dissipative forces. Radial transport can result from different physical forces, e.g. tidal dissipation, solar mass-loss or radiation force, disc effect, gas drag, etc. 

Here, we consider the radial migration of Jupiter, due to the action of the protoplanetary disc and use a Stokes-type dissipative force to mimic the Type-II migration of the planet \citep{bf93,bmfm06} of the form
\begin{equation} \label{Fdissipative}
\bm{F_d}=-C(\bm{v_p}- a \bm{v_c}),
\end{equation}
where $\bm{v_p}$ is the planar velocity component of the planet and $\bm{v_c}$ is the circular velocity at the particular distance of the star. In a first-order approximation, the positive constants $C$ and $a$ are associated with the migration rate in semimajor axis, $\nu$, under $\nu=2C(1-a)$, and the eccentricity damping, $K$, according to $K=\frac{a}{2(1-a)}$. With this simplified dissipative model, we aim to see whether connections between first-order MMRs are possible via the circular family despite the presence of unstable regions associated to second-order MMRs. In other words, we seek for conditions on the migration rate to support the radial transport between different first-order MMRs using the bridge offered by the circular family in the low-eccentricity regime.    
 
In the simulations, we start with (quasi-)circular coplanar orbits for the two bodies. The planet is initially fixed at $a_2(0)=1.0$ au [with $e_2(0)=0, \omega_2(0)=M_2(0)=0^\circ$] and two cases are considered for the initial position of the asteroid: (i) $2/1<n/n'<9/5$ (interior to the $2/1$ MMR) and (ii) $3/2<n/n'<7/5$ (interior to the $3/2$ MMR). The order of the migration rate is given by the formula of \citet{war97} and we consider parameter values in the intervals $10^{-7} \leq \nu \leq 10^{-3}$ yr$^{-1}$ and $0.5\leq K\leq 10$.   

In Fig. \ref{fig2}, we located the asteroid below the 2/1 unstable apocentric branch at $a_1(0)=0.655$, $e_1(0)=0.01$, $\varpi_1(0)=180^\circ$ and $M_1(0)=180^\circ$ (black dot in panel b) with $\theta_{1,2/1}=180^{\circ}$ and we used $\nu=1.25 \times 10^{-4}$ yr$^{-1}$ and $K=4.5$ (rapid migration). The system evolved along the circular family ($e_1\approx 0$), rapidly surpassing higher order resonant orbits. A small yet non-negligible pump of the eccentricities is observed right before the unstable region at 5/3 MMR at $t\approx 0.7$ Myr. The system finally got captured in the 3/2 MMR and followed the stable 3/2 pericentric branch in the CRTBP ($e_1>0$) with $\theta_{1,3/2}=0^{\circ}$ (shown in panel b). The transition between these two first-order MMRs along the circular family is demonstrated by the libration of the respective resonant angles in panel a. 

In Fig. \ref{fig3}, we showcase a similar scenario, where a rapid transition between the 3/2 and 4/3 MMRs takes place. In particular, the asteroid is located below the 3/2 unstable apocentric branch having $a_1(0)=0.78$ and $e_1(0)=0.036$, $\varpi_1(0)=180^\circ$, and $M_1(0)=0^\circ$, with $\theta_{1,3/2}=180^{\circ}$, and we used $\nu\approx 4.2 \times 10^{-4}$ yr$^{-1}$ and $K=1$. After the 4/3 resonance capture, the evolution is bounded by the two branches of the 4/3 MMR in the CRTBP shown in panel (b), but is governed by the pericentric branch with $\theta_{1,4/3}=0^{\circ}$. More precisely, there is a point, right before the unstable segment of the 5/3 MMR located at $a_1/a_2=0.794$ and $e_1=0.0247$, where the two circular families cross each other (one periodic orbit has $\varpi_1=180^\circ$ and $M_1=0^\circ$ and the other has $\varpi_1=180^\circ$ and $M_1=180^\circ$). At this point, the system starts following the stable segment, reaches $e_1=0$, and then gets locked in the 4/3 MMR guided by the 4/3 pericentric branch.

The above examples signify the role played by the migration parameters. When the migration is rapid, higher order MMRs only slightly affect the evolution of the system and captures of the asteroid in adjacent first-order MMRs take place. Unless the orbital migration is rapid, the system will be captured in MMR of higher order, as we will show in the following examples. More precisely, the system has to be located below the 5/3 or 7/5 MMRs [small unstable (red) regions at second-order MMRs on the circular family], so that a capture to the 3/2 or 4/3 MMRs is yielded within a slow migration process.

In Figs.~\ref{fig4} and \ref{fig5}, we provide three illustrations of possible captures in 7/4, 5/3 and 7/5 MMRs. We started from the same periodic orbits of the circular family used in Figs.~\ref{fig2} and \ref{fig3} (i.e. below the unstable apocentric branches of the 2/1 and 3/2 MMRs), but imposed a slower orbital migration with $10^{-7} \leq \nu \leq 10^{-6}$ yr$^{-1}$. The families of periodic orbits in the CRTBP for the 7/4, 5/3 and 7/5 MMRs are displayed in Fig. \ref{fig6}, presented as in Fig. \ref{fig1}.

Regarding the 7/4 MMR, two branches bifurcate from the circular family: one pericentric (stable) with $\theta_{1,7/4}=0^{\circ}$ and one apocentric (unstable) with $\theta_{1,7/4}=180^{\circ}$. The system shown in Fig. \ref{fig4} evolves along the stable pericentric branch of the 7/4 MMR (panel c).

Concerning the 5/3 MMR, as mentioned in Appendix \ref{pert}, at second-order MMRs we have two families bifurcating from each end of the unstable (red) region of the circular family around such MMRs, one being stable and one unstable. In Fig. \ref{fig6}, we present the four families, which are equivalent in pairs, as they differ only in the direction of the perpendicular crossing with the Poincar\'e surface of section. The evolution showcased in panel (c) of Fig. \ref{fig4} is bounded by the two stable apocentric branches with $\theta_{1,5/3}=180^{\circ}$.

Finally, for the second-order 7/5 MMR, Fig. \ref{fig6} shows the four families of periodic orbits in the CRTBP. In Fig. \ref{fig5}, we describe the route followed during the 7/5 resonance capture. The evolution is a bit more complex here since the families of the 7/5 and 4/3 MMRs overlap and both influence the system evolution simultaneously. In panel (a), we provide the evolution of the orbital elements and the resonant angles $\theta_{1,4/3}$ and $\theta_{1,7/5}$; the latter being recomputed with a smaller time-step and output sampling, so that the exact moment of capture in 7/5 MMR is unravelled. In panel (b), we overplot the evolution of the system on the families and illustrate the libration about the two (initially stable) apocentric branches with $\theta_{1,7/5}=180^{\circ}$. The system destabilizes when it reaches the unstable (red) segments at high eccentricity values along these branches. An accumulation of points is evident along the  circular families, as well as along the stable elliptic ones. In order to clarify the path (families) followed during this evolution we inserted the labels A-D. For $0<t<16$ Myr, the system starts from the 3/2 apocentric branch (black dot) and moves along the circular family until point A, which is the intersection point of the two circular families discussed earlier for the 4/3 capture in Fig. \ref{fig3} taking place before the region of instability around the 7/5 MMR. At this point, it starts following the second circular family that is stable and the asteroid reaches an eccentricity value $e_1=0$ for the first time. This behaviour is apparent until $t<27$ Myr, which is marked with the label B. Then, the system gets locked in the 4/3 MMR and evolves along the 4/3 pericentric branch with  $\theta_{1,4/3}=0^{\circ}$ undergoing large amplitude librations until point C, when $t\approx 32.6$ Myr. At $t\approx 36$ Myr, the system reaches point D and gets locked in the 7/5 MMR, allowing the eccentricity of the asteroid to increase. 

\section{Conclusions}\label{con}

Recently, \citet{malzha} and \citet{leili20} analysed the divergence of first-order MMRs for nearly circular direct orbits of a massless body (asteroid) moving in an interior and exterior resonant orbit to Jupiter and measured the libration width on Poincar\'e sections for specific values of the Jacobi constant. They highlighted the displacement of the centres of the distinct pericentric and apocentric resonance zones from the nominal resonance at very low-eccentricity of the test particle and the apparent potential connection of theses zones for adjacent first-order MMRs, which could constitute possible ``bridges'' for radial transport in planetary systems. 

In this work, we reviewed these observations in the context of periodic orbits, emphasizing the continuity of the circular family consisting of resonant and non-resonant orbits in the low-eccentricity regime, except for first-order MMRs where the circular family deviates hyperbolically. When $\mu>0$, the gaps in the circular family associated with the first-order MMRs tear apart the two distinct libration zones, one for the pericentric motion and one for the apocentric motion of the asteroid. Their libration centres deviate from the nominal resonance values \citep[see e.g.][]{col68,gui69,schmiB,schmi72,hadjich84,hadra}. The circular family acts like a bridge that connects orbits of adjacent resonances of different order. Thus, neighbouring first-order MMRs are connected by bridges along the circular family if the second-order (and higher order) MMRs in between are surpassed.

We illustrated by several examples that the circular family can act as a bridge that connects first, second and higher order MMRs. For the Sun-asteroid-Jupiter problem, we showed that, during the migration, due to the interactions of Jupiter with the protoplanetary disc, a system initially evolving below the apocentric branch of a first-order MMR follows the circular family and is captured into the pericentric branch of an adjacent first-order MMR, when the orbital migration is rapid ($10^{-4} \leq \nu \leq 10^{-3}$ yr$^{-1}$). In case of slow migration ($10^{-7} \leq \nu \leq 10^{-6}$ yr$^{-1}$), the system can be captured in a neighbouring higher order MMR present on the circular family. In both cases, the system then evolves along the families of the CRTBP up to possibly high eccentricity of the asteroid.      

The circular family has already been used in the past for multiplanetary systems as a path that drives migrating planets to resonance capture. Regarding the General TBP \citep{hadjbook06}, \citet{av17} showed that the circular family can act as a path that excites the inclination of two planets initially on circular almost coplanar orbits, before they are captured in MMR and follow the spatial elliptic families of the General TBP \citep{vat14}. We believe that radial transport via the circular family can be extended to many small body and planetary system configurations undergoing dissipative effects like tidal dissipation, solar mass loss or radiation force, gas drag, and protoplanetary disc effect. In particular, following the first numerical explorations achieved in this work, the interest of the circular family for bridging MMRs of first or higher order is undeniable.      

\section*{Acknowledgements}
We thank the reviewer, Hanlun Lei, for his astute comments. The work of KIA is co-financed by Greece and the European Union (European Social Fund- ESF) through the Operational Programme \guillemotleft Human Resources Development, Education and Lifelong Learning\guillemotright in the context of the project ``Reinforcement of Postdoctoral Researchers - $2^{\rm{nd}}$ Cycle'' (MIS-5033021), implemented by the State Scholarships Foundation (IKY). The work of ASL was supported by the Fonds de la Recherche Scientifique - FNRS under Grant No. F.4523.20 (DYNAMITE MIS-project).

\section*{Data availability}
No new data were generated or analysed in support of this research.



\bibliographystyle{mnras}
\bibliography{lbib} 

\bsp	
\label{lastpage}

\begin{appendix}
\section{Periodic orbits in the CRTBP}
\label{appendix}
\subsection{The unperturbed case ($\mu=0$)}\label{unpert}
Let us start with the unperturbed case with the mass of Jupiter $\mu=0$. Then, the mass of the Sun $S$ would be equal to 1 and the asteroid $A$, under the gravitational attraction of $S$, would describe a Keplerian orbit around it. This unperturbed orbit of $A$ can be either a circle or an ellipse around $S$ in the inertial frame of reference.  

When viewed in a rotating frame of reference, $Oxy$, with angular velocity $n'>0$ and $S$ being located at its origin, $O$, these two cases of orbits of $A$, with angular velocity $n$, could transition to orbits that are as follows:
\begin{description}
	\item[\textbf{Circular}:] A circular orbit in the inertial frame remains a circular and periodic orbit in the rotating frame. This holds for any radius, $r$, and for both direct (or prograde) (when $n>0$) and retrograde (when $n<0$) orbits. This periodic orbit is symmetric with respect to the $x$-axis and has a period $T=2\pi/(n-1)$, where $n=r^{-3/2}$ and $n'=1$. These are the \emph{periodic orbits of the first kind.} These circular periodic orbits belong to the \emph{circular family} along which a parameter varies, namely the period or the radius (semimajor axis). It is straightforward that we have two distinct circular families, one with direct orbits and one with retrograde orbits. In Fig.~\ref{fig1}, the circular family for $\mu=0$ is found at $e_1=0$ (black line). 
	\end{description}
	\begin{description}
	\item[\textbf{Elliptic}:] The circular periodic orbits of the circular family are not associated in general with an MMR. However, at the circular orbits with $n/n'=p/q$ with $p, q \in Z$ and multiplicity $p-q$, we have the bifurcation of resonant periodic orbits of $A$, which are elliptic in the rotating frame. These are the \emph{periodic orbits of the second kind.} These elliptic periodic orbits belong to families of symmetric elliptic periodic orbits along which the eccentricity of $A$ varies, while the semimajor axis and as a result, the resonance remain constant. Therefore, for each commensurability we have two families (or two \emph{branches}): one corresponding to the location of $A$ at pericentre and one to its location at apocentre, i.e. they differ only in phase, which can be identified by the respective resonant angle. In Fig. \ref{fig1}, we provide examples of such families for $\mu=0$ with black dashed lines. Negative eccentricity values are used when the asteroid is located at apocentre. Finally, we could also have asymmetric elliptic periodic orbits by altering the orientation of the orbit of $A$ with respect to the plane, apart from its eccentricity.
\end{description}

All of the above hold also for the external MMRs, when the asteroid's orbit is exterior to Jupiter's orbit. A global view of the two (direct and retrograde) circular families of periodic orbits together with the bifurcating families of the elliptic 3/2, 2/1 and 3/1 resonant periodic orbits can be found in \citet{hadj88}.

\subsection{The perturbed case ($\mu>0$)}\label{pert}

Let us now move on to the perturbed case for $\mu>0$. In other words, we shall add a sufficiently small perturbation to the motion of $A$ by setting the mass of Jupiter $\mu=0.001$. The symmetric periodic orbits of $\mu=0$ can be continued to $\mu>0$ as nearly circular \citep{birk15} and nearly elliptic \citep{aren63} in the rotating frame. By keeping the mass values constant and by varying another parameter, we can construct the families mentioned above, respectively, i.e. the new ones of nearly circular orbits (circular family) and the new ones of nearly elliptic orbits (CRTBP):
\begin{description}
	\item[\textbf{Nearly circular}:] Given the order of the MMR (bifurcation point of the circular family), we have different behaviour in the evolution and stability of those periodic orbits (coloured curves of Fig. \ref{fig1} at $e_1\approx 0$):
	\begin{itemize}
		\item First-order MMRs: The 2/1, 3/2, 4/3 ... resonant direct circular periodic orbits are not continued as periodic orbits in the rotating frame. \citet{gui69,schmiB,schmi72,hadra} obtained analytically the form of the family when $\mu>0$. It was proved that at these periodic orbits the family of periodic orbits of the first kind (nearly circular) deviates hyperbolically and follows the family of the periodic orbits of second kind (nearly elliptic). 
		\item Second-order MMRs: Around 3/1, 5/3, 7/5 ... resonant direct circular periodic orbits, a small region of instability is created, as the eigenvalues meet at the point $-1$ on the unit circle \citep{hadj82,hadjich84}. According to the Krein - Gelfand - Lidskii theorem, strong stability is preserved if and only if all of the eigenvalues lie on the unit circle and are definite. The eigenvalues move on the unit circle as the period varies along the family of nearly circular orbits and are displaced out of the unit circle on the real axis at second-order resonances. An example of the eigenvalues on the circular family is provided by \citet{hadj92} for the asteroid motion in the 3/1 MMR.
		\item All the rest MMRs: The MMRs having a different order are continued as stable direct symmetric nearly circular periodic orbits in the rotating frame when $\mu>0$.
		\item Non-resonant orbits: These orbits are continued as nearly circular stable orbits.
	\end{itemize}\end{description}
	The behaviour associated with the first-order MMRs does not hold for retrograde orbits and the circular family is continued as a whole when $\mu>0$. Same holds for the linear stability around the second-order MMRs, since the retrograde nearly circular orbits are all stable.
\begin{description}
	\item[\textbf{Nearly elliptic}:] Due to the Poincar\'e - Birkhoff fixed point theorem, out of the infinite set of fixed points, only a finite number survives; half of them being stable and half being unstable. Usually only two survive and they are symmetric with respect to the $x$-axis of the rotating frame. These resonant fixed points bifurcate from the circular family and belong to the families of elliptic periodic orbits (CRTBP) (coloured curves for $e_1\grole 0$ in Fig. \ref{fig1}).
\begin{itemize}
		\item First-order MMRs: Given the discussion above, it is straightforward that the families generated at these MMRs contain two parts: one consisting of nearly circular orbits and one of nearly elliptic orbits. The latter evolves as the eccentricity of $A$ increases in the CRTBP. Examples of the evolution of such families have been shown for the 4/3, 3/2 and 2/1 MMRs and $\mu=0.001$ by \citet{col68,gui69,spis}. In particular, \citet{hadjich84} showed how the pericentric and apocentric branches of first-order MMRs are being connected and also provided surfaces of sections and the linear stability type of the respective periodic orbits by computing the eigenvalues.
		\item Second-order MMRs: From both ends of the small region of instability that appears in the circular family at these MMRs, two families (one stable and one unstable) of elliptic periodic orbits bifurcate to the CRTBP \citep[see][]{hadj93,hadjich84}. 
		\item All the rest MMRs: Likewise each nearly circular periodic orbit is a bifurcation point from which two branches of families of nearly elliptic periodic orbits are generated. These families that belong to the CRTBP differ only in phase and one of them is stable and the other is unstable.
		\item Non-resonant orbits: Such orbits cannot exist, since all the nearly elliptic orbits have to be resonant.
	\end{itemize}
	Examples of planar direct orbits of the CRTBP in internal MMRs can be found in \citet{hadj93as,hadjvoy00,spis}, while planar retrograde orbits in the exterior MMRs were systematically computed by \citet{kovo20p}.
\end{description}

\end{appendix}

\end{document}